# Elastic Properties of Titan Tholins Revealed by Brillouin Spectroscopy

Mark A. Johnson[1], Jose R. Montes-Bojorquez[2], Adis Husić[2], Xinting Yu[2], Wenhao Su[1], Jia-Sen Hu[1], Ryan Blase[3], Edward Patrick[3], Alan Whittington[2,4], and Jin S. Zhang[1*]

[1]Department of Geology and Geophysics, Texas A&M University, College Station, TX, USA.

[2]Department of Physics and Astronomy, University of Texas at San Antonio, San Antonio, TX, USA.

[3]Southwest Research Institute, San Antonio, TX, USA.

[4]Department of Earth and Planetary Sciences, University of Texas at San Antonio, San Antonio, TX, USA.

*Corresponding author: Jin S. Zhang (jinzhang@tamu.edu)

## Key Points:

- First direct measurements of tholins' complete set of elastic properties offer key constraints for Dragonfly seismic interpretations.
- The high shear rigidity of tholins supports their crucial role in Titan dune stability.
- Prolonged atmospheric exposure causes persistent stiffening of tholins, emphasizing the need for careful sample handling and storage.

## Abstract

Titan tholins, complex organic solids synthesized by simulating photochemical processes in Titan's atmosphere, are widely regarded as analogs to organic surface materials on the Saturnian moon. Their mechanical properties are critical for understanding the evolution of Titan's geomorphology and crustal layering. Here, we report the first direct experimental determination of the complete set of aggregate elastic moduli of tholins using Brillouin spectroscopy. Measurements conducted on tholin powder samples yield a bulk modulus of 11(2) GPa and a shear modulus of 3.9(5) GPa, from which we derive a Young's modulus of 11(1) GPa and a Poisson's ratio of 0.34(3). Thin-film samples synthesized under slightly different conditions yield similar results. Compared with other Titan-relevant organic compounds, tholins exhibit the highest elastic stiffness measured to date, exceeding even water ice in both bulk and shear modulus. These results support the interpretation that complex organics contribute significantly to the mechanical stability and transport behavior of Titan's equatorial dune sediments. Measurements following prolonged atmospheric exposure reveal persistent stiffening of tholins that remains after vacuum treatment, likely associated with small chemical modification and physical reorganization, highlighting the importance of controlled sample handling for future laboratory investigations. This study also demonstrates the effectiveness of Brillouin spectroscopy as a non-destructive technique for characterizing planetary organic materials and provides new constraints relevant to future seismic and geophysical investigations by the Dragonfly mission.

## Plain Language Summary

Titan, Saturn's largest moon, has a surface covered with complex organic materials. These materials are thought to form in Titan's atmosphere and settle onto the surface, where they influence how the landscape evolves. However, we still know very little about how strong or flexible these materials are, which makes it difficult to understand processes such as dune formation, erosion, and surface activity. In this study, we produced analogs of these organics in the lab, called "tholins", and measured how they respond to stress using a technique called Brillouin spectroscopy, which allows us to determine how fast sound waves travel through a material. From these measurements, we calculated how stiff and compressible tholins are. We found that tholins are relatively stiff compared to other materials found on Titan, meaning they are more difficult to deform. This new information helps scientists better understand how Titan's surface forms and changes over time, and will also support future missions, such as Dragonfly, that aim to explore Titan's unique environment.

## 1 Introduction

Titan, Saturn's largest moon, has long attracted interest from planetary scientists due to its Earth-like characteristics, including a dense atmosphere and a geologically active surface. Its atmosphere is composed primarily of ~98% diatomic nitrogen ($N_2$) and on average ~2% methane ($CH_4$), with methane concentrations ranging from 1.48% in the lower stratosphere to 5.65% near the surface (Niemann et al., 2010). In addition, trace amounts of hydrocarbons, nitriles, water, carbon dioxide, carbon monoxide, and diatomic hydrogen contribute to Titan's complex atmospheric chemistry and surface dynamics (Nixon 2024). These species are formed through interactions between atmospheric molecules and energetic inputs such as charged particles from Saturn's magnetosphere, cosmic rays, and solar irradiation. As these reactions proceed, large organic molecules undergo further reaction with atmospheric species to produce negatively charged massive molecules, which subsequently aggregate into solid organic materials (Waite et al., 2007). These complex organic particles remain suspended in the atmosphere for extended periods, forming Titan's characteristic dense haze layers, before settling onto the surface. Because the chemical processes that produce these complex organics are not unique to Titan, they are expected to be widespread throughout the Solar System. Evidence of complex organics has been observed on the surfaces of other planets' moons, in the atmospheres of gas giants, and on dwarf planets, comets, asteroids, trans-Neptunian objects (Boehnhardt et al., 2004; Cruikshank et al., 2019a), and even within the interstellar medium (Sagan and Khare, 1979).

Following their deposition on Titan's surface, these complex organic particles play a key role in various surface geologic processes on Titan, including the formation of dunes (Yu et al., 2017). Longitudinal dunes are widespread on Titan, covering 17% of its surface, mainly near the equator (Lorenz et al., 2006; Lopes et al., 2020). Infrared spectroscopic observations acquired by the Visible and Infrared Mapping Spectrometer (VIMS) aboard the Cassini-Huygens spacecraft indicate that these dunes are predominantly composed of organic material (e.g., Soderblom et al., 2007). The organic-rich composition of the dunes suggests they likely originate from the complex organic haze particles produced in Titan's atmosphere (Barnes et al., 2008; Lopes et al., 2010). On Earth, such dunes can be formed either through bidirectional winds acting on loose sand, or unidirectional winds transporting sand over stabilized sediment. Rubin & Hesp (2009) proposed the latter mechanism for Titan, in which surficially cohesive materials are crucial to dune formation. Atomic force microscopy (AFM) and contact angle goniometry measurements on Titan tholins supported this interpretation (Yu et al., 2017). In this framework, complex organics and other particles accumulate and become stabilized, forming surface regions resistant to aeolian erosion that function as obstacles to wind flow. As wind passes over these obstacles, its velocity decreases, promoting sediment deposition downwind and leading to the elongation of dunes to hundreds of kilometers in length over time (Rubin & Hesp, 2009). Through continued transport and accumulation, Titan's dunes have grown into a substantial global reservoir of organic materials. With an estimated total volume of 2–8 × $10^5$ $km^3$, they represent one of the largest surface stores of organics on Titan (Lorenz et al., 2008).

The presence of organic fluvial systems, impact craters, and active geology indicates that the atmospheric haze particles may not only accumulate on the surface but also react and even be incorporated into Titan's deep icy shell (Neish et al., 2024). The mechanical properties, including bulk modulus ($K_S$), shear modulus ($G$), Young's modulus ($E$), and Poisson's ratio ($v$) of the complex organics on Titan, are important for understanding the surface processes outlined above, and subsurface and crustal dynamics. In the context of the upcoming Dragonfly mission, seismic measurements will provide new constraints on Titan's near-surface sedimentary structures and deep crustal layering. Interpreting these observations requires the elastic moduli of relevant material constituents, especially Titan's complex organics.

Since the first successful laboratory synthesis of Titan's haze analogs by Sagan and Khare (1979), who named the material "tholins", tholins have been studied extensively as analogs of Titan's organic aerosols and surface deposits for decades (e.g. Sarker et al., 2003; Yu et al., 2020). The properties of tholins are known to vary with synthesis conditions; gas composition, pressure, temperature, and the type and energy of the irradiation source can affect various chemical and physical properties of tholins (Cable et al., 2012; Imanaka et al., 2004; Li et al., 2022; McDonald et al., 1994; Sagan & Khare, 1979; Sciamma-O'Brien et al., 2023). Previous investigations have characterized their chemical composition (Carrasco et al., 2009; Imanaka et al., 2004; McDonald et al., 1994; Morrison et al., 2016; Sagan & Khare, 1979), optical properties (Imanaka et al., 2004; Khare et al. 1984; Mahjoub et al., 2012), morphology (Carrasco et al., 2009), surface interactions (He & Smith, 2014; Li et al., 2022; Yu et al., 2017, 2020), electrical properties (Lethullier et al., 2018), and thermal stability (He & Smith, 2014; Nna-Mvondo et al., 2013) using a range of experimental techniques. In terms of the mechanical properties, there are only a handful of available studies: Using nanoindentation combined with AFM, Yu et al. (2017) measured an interparticle force of 0.8(6) μN and derived a $E$ of 3.0 GPa. In a subsequent study, Yu et al. (2018) utilized nanoindentation and obtained an updated value for $E$ of 10.4(5) GPa, along with nanoindentation hardness ($H$) and fracture toughness ($K_c$) values of 0.53(3) GPa and 0.035(7) MPa·m$^{1/2}$, respectively. The Young's modulus value of 10.4(5) GPa reported in Yu et al. (2018) is considered more reliable than the value reported in Yu et al. (2017) because the tip geometry was directly calibrated against a reference material with a known elastic modulus. These studies provide valuable constraints on the local mechanical behavior of tholins at small scales. However, the full set of aggregate elastic properties, particularly $K_S$, $G$, and $v$, has never been directly measured. This lack of data on a geologically important material limits our ability to quantitatively assess the mechanical properties of Titan's crust, as well as interpret future seismic data obtained from the upcoming Dragonfly mission (Lorenz et al., 2018). In this study, we employ Brillouin spectroscopy to directly measure the longitudinal ($V_P$) and shear ($V_S$) acoustic velocities of two types of tholin samples, enabling the first complete determination of their elastic properties at ambient pressure–temperature conditions. We then evaluate the significance of these results and their implications for planetary processes.

# 2 Methods

## 2.1 Sample synthesis, characterization and preparation

In this work, we use the Tholinator setup at the Southwest Research Institute's Space Environment Simulation Lab to generate tholin samples. A detailed description of the experimental setup and characterization of the tholin samples can be found in Husić et al. (2026). In this system, a constant gas flow is exposed to a cold plasma generated using a radio-frequency (RF) discharge operating at 13.56(1) Hz in a glass reaction vessel, inducing reactions between simple gas molecules that ultimately produce tholins. The Tholinator setup can produce tholins in two forms: thin films and powders composed of spherical particles. The thin film samples represent the first stage of tholin production, where tholins are deposited on the inner walls of the reaction vessel as a result of the wall effect (Cable et al., 2012). Upon complete coating of the inner walls with a thin film of tholins, the wall effect ceases, and the system transitions to a second production mode where spherical particles nucleate and grow freely independent of the wall surface (Husić et al., 2026). These spherical particles more closely resemble the aerosol morphology expected in Titan's atmosphere, as their formation is not affected by the wall effect. Due to the known susceptibility of tholins to degradation in open air (Cable et al., 2012), all samples were collected and then stored under vacuum, with minimal exposure to the atmosphere, until measurement.

The powder sample used in this work was generated using a gas mixture of 95% $N_2$ and 5% $CH_4$ at a pressure of 1 Torr. A flow rate of 7 standard cubic centimeters per minute was maintained to ensure the pressure remained constant throughout the experiment. The RF plasma was generated with a forward power of 36 W. The synthesis experiment was conducted for ~500 hours and yielded ~1 g of tholin powder. The narrow particle-size distribution suggests relatively uniform particle morphology, although chemical heterogeneity at the particle scale cannot be completely ruled out (Husić et al., 2026). The bulk chemical composition of the tholin sample was characterized using high-resolution mass spectrometry and will be reported in a forthcoming companion study.

The thin film sample used in this work was generated using the same setup, but with a gas mixture of 90% $N_2$ and 10% $CH_4$, a pressure of 1 Torr, a flow rate of 6 standard cubic centimeters per minute, a forward power of 30 W, and a reaction time of 6 hours. A higher methane concentration was used for this sample to ensure uniform deposition of the sample. To synthesize the thin film sample, a 6 mm square, 25(7) μm-thick a-cut sapphire platelet was mounted on a silica glass slide and cleaned by UV-ozone treatment immediately before deposition to remove organic contaminants. The mounted substrate was then placed inside the glass reaction vessel prior to synthesis. A thin sapphire substrate (25 μm thick) was utilized for two primary reasons. First, sapphire can be fabricated at sufficiently small thicknesses, reducing the substrate scattering signal, which is important because the tholin film thickness and corresponding scattering volume are expected to be exceedingly small along the laser path. Second, the shear acoustic velocities ($V_s$) of sapphire are significantly higher than those expected for typical organic compounds (Zhang et al., 2024), preventing overlap or interference between the substrate and sample Brillouin signals. In contrast, other common substrates, such as glass, exhibit $V_s$ (~3000-3800 m/s) that may overlap with the longitudinal modes ($V_p$) of tholins, complicating spectral interpretation. During the experiment, tholins were deposited on both the inner walls of the glass tube and on the surface of the sapphire platelet.

The thickness of the thin film sample was determined by spectroscopic ellipsometry (SE) using a J.A. Woollam M2000-DI variable-angle spectroscopic ellipsometer operating in a MBraun dry nitrogen glovebox (<2 ppm of water vapor and oxygen) at the Planetary Material Characterization Facility (PMCHEF) at the University of Texas at San Antonio, and data were analyzed in the software *CompleteEASE* (J.A. Woollam Co.). Film thickness was estimated using the mounting glass slide as a witness substrate. This approach was preferred over direct measurements on the sapphire-supported film because the transparent, optically anisotropic a-cut sapphire substrate introduces additional optical contributions that make thickness extraction by ellipsometry unreliable without modifying the sample. While direct measurement may be possible after backside/surface modification or by destructive cross-sectional methods, these approaches were avoided to preserve the sample for Brillouin spectroscopy. The reported film thickness, 930(70) nm, corresponds to the average thickness measured across the surrounding witness regions and assumes that the deposition thickness did not vary significantly between the sapphire substrate and the neighboring glass-slide region.

The tholin powder sample was loaded into a diamond anvil cell (DAC) with 450 μm culet diamonds and a 20 μm thick rhenium gasket. A cylindrical sample chamber (~250 μm in diameter) was drilled into the gasket, which was then sandwiched between the two diamond anvils. The DAC was gently compressed to reduce pore space between individual tholin particles until grain boundaries were no longer observable by visual inspection using a stereomicroscope, producing an optically homogeneous solid. Although residual microporosity below the spatial resolution of optical microscopy cannot be excluded, the absence of visible pores and grain boundaries following compression indicates that large voids and interfaces capable of producing significant optical scattering were largely eliminated and are thus unlikely to affect the Brillouin scattering measurements. To avoid over-compression, the sample was closely monitored under the stereomicroscope during loading, and pressurization was immediately stopped upon the disappearance of grain boundaries. Given the relatively low $E$ (~10 GPa) and expected yield strength on the order of tens of MPa (Yu et al., 2018), only modest pressures (<1 GPa) are sufficient to eliminate pore space and produce a dense, optically homogeneous sample. Under such low-pressure conditions, any physical changes to the tholin particles are expected to be limited to minor elastic compaction and pore collapse, with minimal irreversible structural modification. This step minimizes the elastic scattering at particle boundaries and enhances sample transparency, which is critical for forward scattering Brillouin measurements, where signal intensity depends on the amount of light transmitted through the sample. To directly quantify the pressure conditions of the tholin pellets during powder compaction and after decompression, we conducted ruby fluorescence experiments on 8 ruby spheres loaded with the powder tholin samples in the same sample chamber (Figure S1). As expected, the average pressure during compression was only 0.87(6) GPa. After decompression, a small residual pressure of 0.19(3) GPa was obtained, which may reflect a combination of residual internal stresses retained during irreversible sample compaction and systematic uncertainties of ruby pressure scale associated with variable Cr concentrations among the ruby spheres. The thickness of the sample pellet is estimated to be ~10 μm following compression due to the collapse of interparticle voids. Following compression, the DAC was decompressed prior to data acquisition to ensure measurements were conducted under ambient pressure conditions. The thin film tholin sample was measured directly on the sapphire platelet without any additional preparation.

The average size of the tholin particles before compression is ~1 μm, following a Gaussian distribution (Husić et al. 2026). Following compression and subsequent decompression, the sample remains transparent and grain boundaries do not reappear, suggesting irreversible particle consolidation. However, we cannot distinguish whether this is caused by the high interparticle forces described in Yu et al. (2017), chemical diffusion across grain boundaries during compression, or a combination of both. The thin-film sample is also expected to have a relatively compact microstructure, as due to the wall effect, particles in the thin-film can grow over 100 μm in width, resulting in a more compact structure with less porosity. SEM images of both the initial powder and thin-film samples can be found in Husić et al. (2026).

Additional Brillouin spectroscopy experiments were performed on the same tholin powder sample following one week of exposure to ambient atmospheric conditions, and subsequently after one day of vacuum pumping, to evaluate the effects of prolonged atmospheric exposure on tholin stiffness and whether any detected changes are persistent after vacuum treatment. Relative humidity was monitored intermittently using a HOBO humidity/temperature data logger, which recorded values ranging from ~30% to ~80% throughout the week. Fourier Transform Infrared (FTIR) spectroscopy measurements were conducted to evaluate any chemical alterations resulting from atmospheric exposure. Measurements were conducted on the compressed powder tholin pellets prepared using the procedure as for the Brillouin spectroscopy experiments. Spectra were collected using a Bruker Hyperion II FTIR microscope equipped with an MCT-A detector. Spectra were acquired from the pristine sample immediately after removal from a vacuum desiccator, after one day of ambient atmospheric exposure, and after prolonged (>1 week) exposure. Before each measurement, the sample was placed in the vacuum desiccator for at least 2 hours prior to each measurement to minimize the contribution of adsorbed water and focus on the persistent changes to the tholins.

Early signs of atmospheric exposure were readily identifiable in the powder sample through several observable changes. With prolonged exposure, the powder agglomerated into larger, more irregular particles and became slightly darker in color. Upon compression, the exposed samples also exhibited greater heterogeneity under optical microscopy. In addition, the exposed powder adhered more readily to the needle used for sample transfer, indicating increased surface adhesion following atmospheric exposure. Visual identification of these exposure-induced changes was more difficult for the thin-film sample because of its small thickness. Nevertheless, changes were evident from the Brillouin spectra relative to those of the freshly prepared film.

### 2.2 Brillouin scattering experiments and data analysis

Brillouin scattering measurements were conducted in the Geospectroscopy Laboratory at Texas A&M University. Experiments were performed with a 300 mW 532 nm single-mode diode-pumped solid-state laser in conjunction with a TFP1 6-pass tandem Fabry-Pérot interferometer. Details of the experimental setup and calibration process are described in previous literature (Johnson et al., 2026; Zhang et al., 2015; 2025). A 50° symmetric forward scattering geometry was used, with the angle precisely calibrated to 50.44°. Due to the ultra-low elastic moduli, and thus low acoustic velocities of tholins, large interferometer plate spacings (15 and 20 mm) were employed in all experiments. To assess possible effects of sample heterogeneity and spatial variations in stress, Brillouin spectroscopy measurements were conducted at two different locations on each of the powder and thin film samples, with a laser spot size of ~25 μm. Because tholins exhibit broad absorption extending from the UV into the visible wavelengths, including a non-negligible absorption tail at 532 nm (Imanaka et al., 2004), the incident laser power was attenuated to ~750 μW using a linear wire grid polarizer to mitigate laser-induced damage to tholin samples, which were examined under stereomicroscopes after each experiment. Consequently, acquisition time of 12–24 hours per measurement was required to achieve sufficient signal-to-noise ratios. During these long acquisitions, small changes in the interferometer free spectral range were accounted for by averaging the calibration obtained before and after each measurement. The resulting uncertainty in acoustic velocities was 4–8 m/s, substantially smaller than both the Brillouin peak widths and the uncertainties in velocities derived from the peak positions. The asymmetric shape of the central Rayleigh peak, a known instrumental artifact of the TFP-1 interferometer, does not affect the determination of the Brillouin peak positions.

The spectra reported in this study were collected not only at different positions on the samples but also at different azimuthal angles within the vertical plane. We observed no systematic angular dependence in either the Brillouin peak shapes or the measured acoustic velocities. Thus, although texture-induced elastic anisotropy cannot be completely ruled out, we treat the tholins as approximately isotropic, consistent with their amorphous nature; accordingly, only a single shear acoustic mode is resolved in addition to the longitudinal mode.

Brillouin scattering experiments measure the frequency shifts ($\Delta\omega$) of light resulting from its interaction with acoustic phonons in a material. Both Stokes and anti-Stokes components result from the exchange of energy between incident photons and thermally excited acoustic modes within the sample. Under symmetric forward scattering geometry, the magnitude of $\Delta\omega$ is linearly proportional to the acoustic wave velocities ($V$) of the sample, as given by Equation (1):

$$2V \sin\left(\frac{\theta}{2}\right) = \lambda\Delta\omega \tag{1}$$

where $\theta$ is the scattering angle and $\lambda$ is the incident wavelength. Longitudinal and transverse modes correspond to $V_p$ and $V_s$, respectively.

A low-pass Savitzky-Golay filter was applied to each Brillouin spectrum to reduce noise for peak fitting. The choice between the Savitzky-Golay filter and other filters (e.g. a conventional low-pass moving average filter) has a negligible effect on the derived sound velocities (Figure S3). Peak positions were then manually fitted using the software *Fityk* (Wojdyr, 2010). Because the Brillouin peaks across all spectra were sufficiently strong and clearly distinguishable from the broad, slowly varying fluorescence background, no background subtraction was applied prior to peak fitting. Aggregate $K_S$ and $G$ can be calculated using Equations (2) and (3):

$$K_s = \rho \left( V_p^2 - \frac{4}{3} V_s^2 \right) \quad (2)$$

$$G = \rho V_s^2 \quad (3)$$

where $\rho$ is the density of the tholin samples, determined as 1.4(2) g/cc from the helium pycnometry measurements (Husic et al. 2026). $E$ and $v$ can be derived from $K_S$ and $G$ by Equations (4) and (5):

$$E = \frac{9K_sG}{3K_s + G} \quad (4)$$

$$v = \frac{3K_s - 2G}{2(3K_s + G)} \quad (5)$$

## 3 Results and Discussions

Spectra collected from multiple locations of different tholin samples consistently exhibit broad, yet well-defined Brillouin peaks with highly reproducible peak positions (Figure 1), demonstrating the robustness of the measurements. The powder sample was composed of particles ~1 µm in size prior to compression (Husić et al. 2026). Following compression, individual grain boundaries were no longer optically distinguishable, and the sample appeared homogeneous at the spatial resolution of optical microscopy. Because the laser spot (~25 µm in diameter) encompasses numerous particles, the measured Brillouin signal represents an effective aggregate response of the consolidated powder. Although the observed Brillouin peaks are substantially broader than those typically observed in single-crystal silicate minerals, this peak broadening most likely reflects strong acoustic attenuation associated with viscoelastic damping (i.e., internal friction and relaxation processes, (Palombo and Fioretto, 2019), as Brillouin peak width is known to reflect viscous dissipation in materials (Antonacci et al., 2020; Bailey et al., 2020). In contrast, inhomogeneous peak broadening resulting from the chemical and structural heterogeneities of tholins is unlikely to be the primary contributor to the observed large peak width, given the similarity of Brillouin peak shapes and positions measured at multiple locations on each sample. However, heterogeneity at length scales significantly smaller than the area probed by Brillouin measurements cannot be completely ruled out, as measurements at different locations may sample similar distributions of sub-resolution heterogeneities. Likewise, stress-induced broadening is likely negligible given the small residual pressure (~0.19 GPa) and deviatoric stress inferred from ruby fluorescence measurements on multiple ruby spheres embedded within the compressed powder pellet (Figure S1). Although the thin-film sample retains a different microstructure with larger lateral features than the consolidated powder, the consistent P- and S-wave velocities and derived elastic properties obtained from both sample types indicates that neither microstructural differences, gentle DAC compression, nor minor variations in synthesis conditions (e.g. wall effect for the thin film sample and differences in the initial gas mixture composition), have a significant influence on the measured elastic properties of tholins. The slightly broader and less symmetric Brillouin peaks observed in the thin-film sample likely result from its distinct microstructure, with larger, densely packed particles produced during deposition (the wall effect), leading to greater heterogeneity and more internal interfaces than in the optically transparent consolidated powder pellets. Importantly, these differences affect peak shape but do not produce measurable differences in acoustic velocities.

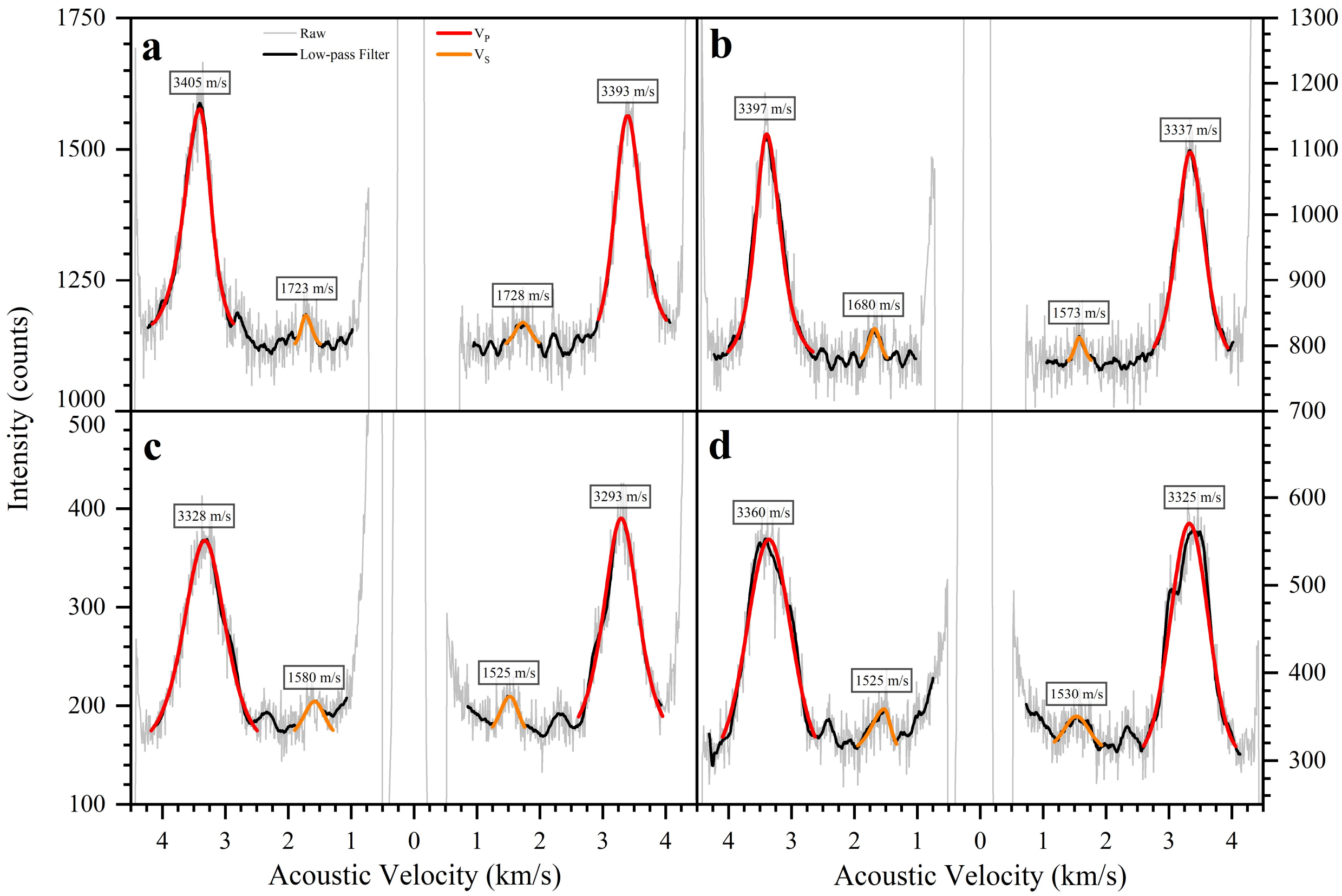

**Figure 1:** Brillouin spectra collected from the two tholin samples: (**a, b.**) spectra acquired at different locations on the tholin powder platelet in the DAC. (**c, d.**) spectra acquired at different locations on the tholin thin film. The thin gray curves represent the raw spectra, whereas the thick black curves show the spectra after smoothing with a low-pass filter. The red and orange curves indicate the $V_p$ and $V_s$ peaks, respectively. The scattering signals are symmetric about the central Rayleigh peak, which has been attenuated to preserve the signal, with the left and right sets of peaks on each subpanel corresponding to anti-Stokes and Stokes scattering, respectively.

The $K_S$ and $G$ of the tholin powder were determined to be 11(2) GPa and 3.9(6) GPa, respectively, corresponding to $E$ = 11(1) GPa and $\nu$ = 0.34(3). These values are consistent with the $E$ = 11.1(6) GPa reported by Husić et al. (2026) for thin-film samples produced under the same conditions using a 95% $N_2$ / 5% $CH_4$ gas mixture in the Tholinator setup. The elastic properties of the tholin film synthesized in this study under slightly different conditions (90% $N_2$ / 10% $CH_4$) are similar to those of the powder sample within experimental uncertainties, yielding $K_S$ = 11(2) GPa, $G$ = 3.3(5) GPa, $E$ = 9(1) GPa, and $\nu$ = 0.36(3). Our measurements also broadly agree with the $E$ = 10.4(5) GPa reported by Yu et al. (2018) for a tholin sample synthesized using a 95% $N_2$ / 5% $CH_4$ mixture in a different laboratory (He et al., 2017). The elastic properties of tholins determined in this study are also compared with other compounds relevant to Titan (Table 1). Among all Titan-relevant organic materials studied thus far, tholins exhibit the highest elastic moduli, indicating relatively high stiffness. In particular, the $K_S$ and $G$ of tholins are slightly higher than those of pure water ice $I_h$ (by 23% and 13%, respectively) and methane clathrates (by 36% and 26%, respectively), suggesting relatively small seismic contrasts between near-surface organic sediments enriched in haze particles and Titan's icy crust (Gammon et al. 1983). Such weak contrasts may present challenges for resolving subsurface crustal layering using seismic observations acquired by the Dragonfly mission. Although typical organic molecular crystals are dominated primarily by relatively weak van der Waals interactions, and water ice by hydrogen bonding, tholins are characterized by complex heteropolymer structures (Sagan & Khare, 1979). This irregular cross-linking likely contributes significantly to tholins' increased stiffness relative to other compounds.

| **Material** | **$K_S$ (GPa)** | **$G$ (GPa)** | **$E$ (GPa)** | **$\nu$** | **Method** | **Reference** |
|---|---|---|---|---|---|---|
| **Tholin powder** | 11(2) | 3.9(6) | 11(1) | 0.34(3) | Brillouin scattering | This study |
| **Tholin film** | 11(2) | 3.3(5) | 9(1) | 0.36(3) | Brillouin scattering | This study |
| **Ice $I_h$** | 8.92(2) | 3.46(6) | 9.20(4) | 0.3282(8) | Brillouin scattering† | Gammon et al. (1983) |
| **Biphenyl** | 2.6(7) | 0.9(2) | 2.4(6) | 0.35 | Nanoindentation* | Yu et al. (2018) |
| **Naphthalene** | 9(1) | 3.1(4) | 8(1) | 0.35 | | |
| **Coronene** | 2.5(9) | 0.8(3) | 2.3(9) | 0.35 | | |
| **Phenanthrene** | 7(1) | 2.2(3) | 5.9(9) | 0.35 | | |
| **Methane clathrate hydrate** | 8.1 | 3.1 | 8.2 | 0.33 | Brillouin scattering† | Shimizu et al. (2002) |
| **Hydroquinone** | 8.7(2) | 3.4(3) | 9.0(6) | 0.33(1) | Brillouin scattering | Zhang et al. (2025) |
| **Benzene** | 4.60 | 1.76 | 4.68 | 0.33 | Ultrasonic interferometry† | Heseltine et al. (1964) |

**Table 1**. Comparison of elastic properties for Titan-relevant materials measured at or near ambient conditions. Reported values of $K_S$, $G$, $E$, and $v$ are listed together with the corresponding experimental methods. †Elastic properties calculated from reported single-crystal elastic moduli. *$K_S$ and $G$ calculated assuming $v = 0.35$ as a representative value for organic materials in the absence of independent measurements.

As stated in the methods section, all tholin samples were stored under vacuum in a desiccator prior to measurement. During data acquisition, the powder sample experienced negligible exposure to air for approximately 24 hours, as it was contained within the diamonds and the gasket of the DAC, while the tholin film sample on the sapphire substrate was directly exposed to the atmosphere throughout the experiments. Tholin samples are known to be susceptible to degradation under humid atmospheric conditions; however, no discernible changes were noticed in the Brillouin spectra between the initial and final measurements of the same sample, suggesting that atmospheric exposure did not significantly affect the bulk elastic properties of the tholin samples within 24 hours. This is further supported by our FTIR measurements, which upon qualitative observation, show no detectable differences after 24 hours of atmospheric exposure (Figure S2). A forthcoming companion study will quantitatively characterize possible chemical changes resulting from atmospheric exposure.

Following these measurements, the powder sample was subsequently exposed to the atmosphere for one week, and we collected an additional Brillouin spectrum (Figure 2). Clear shifts toward higher frequencies in both the longitudinal and transverse modes were observed. This indicates that prolonged atmospheric exposure leads to an increase in stiffness, assuming density did not change, with the exposed sample exhibiting higher elastic moduli of $K_S$ = 13(2) GPa, $G$ = 4.7(7) GPa, $E$ = 13(2) GPa, and $v$ = 0.34(3). This corresponds to an 18% and 21% increase in $K_S$ and G, respectively, indicating broadly comparable stiffening of the volumetric and shear responses during atmospheric exposure, with no clear preferential effect on either deformation mode. The same exposed sample was subsequently positioned in a desiccator under active pumping for one day and then measured again using Brillouin spectroscopy. The observed shift persisted after vacuum treatment, indicating that the stiffening cannot be explained solely by readily reversible water absorption or by differences in the water content during measurements. The P- and S- wave peaks in the exposed sample are much narrower than those of the pristine powder and thin-film sample. The origin of this peak narrowing cannot be uniquely determined by Brillouin scattering measurements alone. It may reflect changes in the intrinsic mechanical or structural properties of the tholin sample during atmospheric exposure, such as reduced acoustic attenuation. Alternatively, chemical alteration may further reduce the already limited residual microporosity, or trace amounts of adsorbed water may redistribute within the micropores, thereby decreasing scattering at particle interfaces and contributing to the observed peak narrowing.

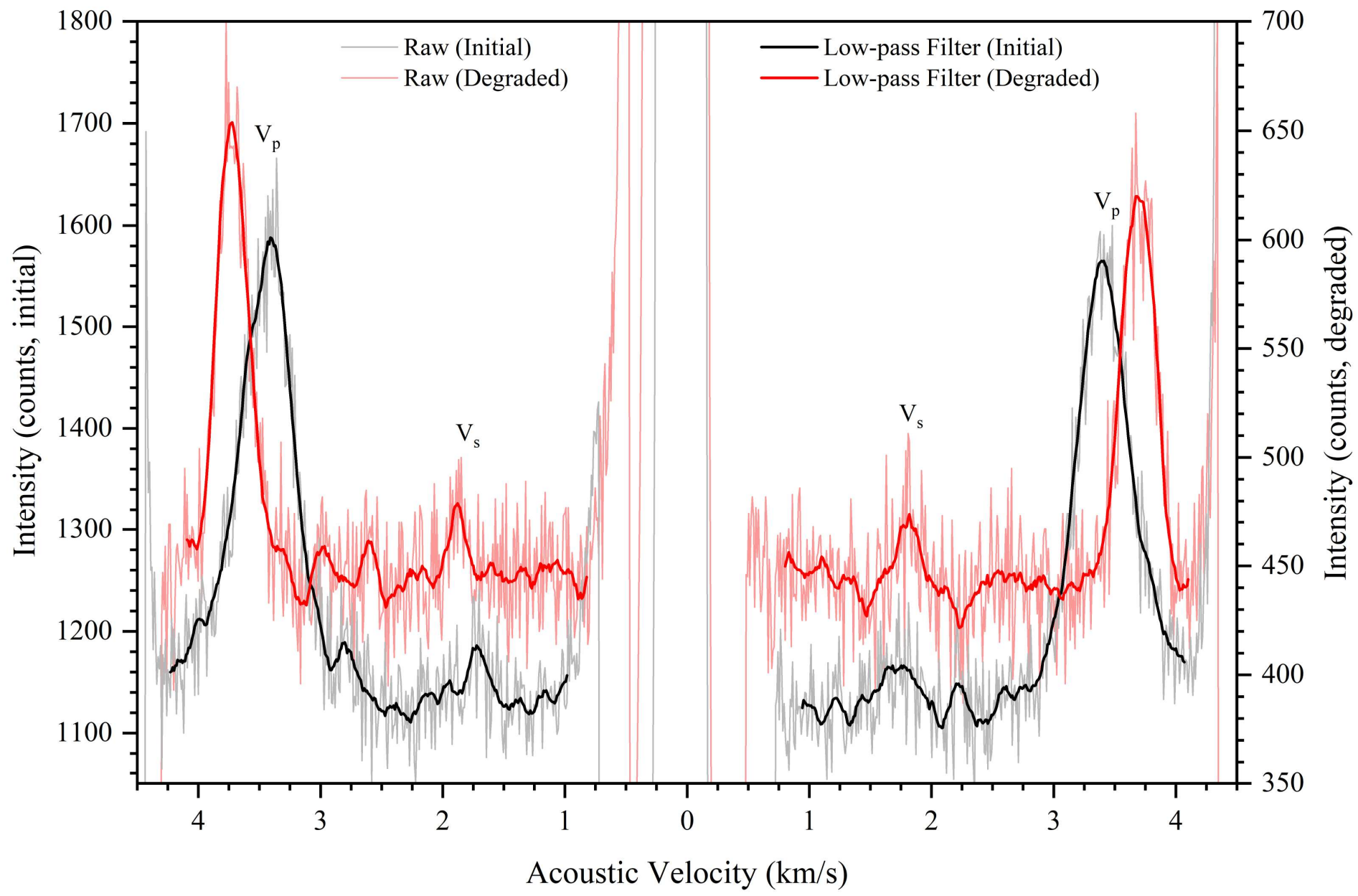

**Figure 2:** Brillouin spectrum of the same tholin powder sample before and after prolonged atmospheric exposure. The lower black and gray curves correspond to the initial spectrum shown in Figure 1a, acquired prior to atmospheric exposure, whereas the upper, red and pink curves represent the spectrum collected after atmospheric exposure. The thin gray and pink curves show the raw spectra, while the thick black and red curves represent spectra smoothed with a low-pass filter.

To evaluate whether atmospheric exposure produced detectable chemical changes or differences in water-related absorption, FTIR spectra were collected from the powder sample immediately after removal from the vacuum desiccator and after short-term (1 day) and prolonged (>1 week) exposure. FTIR spectra collected on the pristine powder sample immediately after preparation and after one day of exposure to ambient conditions show no noticeable differences, indicating that no significant chemical modification was detected within the first 24 hours (Figure S2). Despite no discernible changes in the O-H/N-H region (2900-3800 cm$^{-1}$), the spectra of the sample subjected to prolonged atmospheric exposure exhibit spectral changes in the 1500-1700 $cm^{-1}$ region. Although the overlapping absorption bands in this region preclude assignment of specific functional groups (Gautier et al., 2012), these changes suggest minor chemical modification associated with atmospheric aging. The similar magnitude of the absorption coefficients also does not support substantial bulk densification and therefore supports the use of the same density for calculating the elastic moduli of the pristine and exposed samples.

Taken together, our results suggest that the stiffening following atmospheric exposure may reflect a combination of small chemical and physical changes. Because the elastic moduli were calculated assuming the same density for the pristine and exposed samples, the derived stiffening primarily reflects the increase in acoustic velocities, although smaller density changes cannot be excluded. The persistence of the Brillouin shifts after vacuum treatment, together with the similar water-related absorption among the samples, suggests that the response cannot be explained solely by differences in readily absorbed water. However, water may still contribute through redistribution within residual microporosity or changes in its binding environment. Capillary forces associated with water in small pores could alter interparticle contacts or contribute to local compaction (Scherer, 1988). These physical effects could produce a stiffer elastic response without major changes in the FTIR spectra. Minor chemical modification may also contribute, for example, oxidation of reactive functional groups could introduce oxygen-bearing species and strengthen inter- or intramolecular interactions (Nuevo et al., 2022). The relative contributions of these processes cannot be resolved from the current measurements. Atmospheric alteration and the reversibility of the associated changes under controlled environmental conditions are the subject of an ongoing study.

Experiments on tholins also provide several insights for optimizing Brillouin spectroscopy experiments on highly absorbing and fluorescent organic compounds. First, we demonstrate that, with careful alignment of the TFP-1 interferometer, long-term stability (up to 24 hours) can be achieved even at exceedingly large plate spacings (up to 20 mm). Such large plate spacings are necessary to resolve Brillouin peaks in soft organic compounds with low acoustic velocities. Second, tholins exhibit a broad fluorescence spectrum with a peak near 471 nm, relatively close to the 532 nm laser wavelength used in this study (Hodyss et al., 2004). Fortunately, although the fluorescence resulted in an elevated background, the Brillouin peaks remained resolvable. Finally, strong optical absorption and the resulting opacity present another experimental challenge. Thicker samples reduce light transmission, whereas thinner samples are more susceptible to alteration during atmospheric exposure. We found an optimal powder sample thickness of approximately 20 μm that provides sufficient transparency for reliable measurements, while maintaining stability over 24 hours of ambient exposure. Another relevant issue is the optimization of laser power. Higher laser power can cause thermal degradation of organic compounds, such as tholins (He & Smith, 2014), whereas excessively low laser power results in weak signals and impractically long acquisition time. Systematic testing showed that laser powers exceeding ~850 μW produce localized "bleaching" at the measurement spot, manifested by increased transparency and lighter coloration, indicating sample alteration (Figure 3b). By limiting laser power to below ~750 μW, such damage could be effectively avoided.

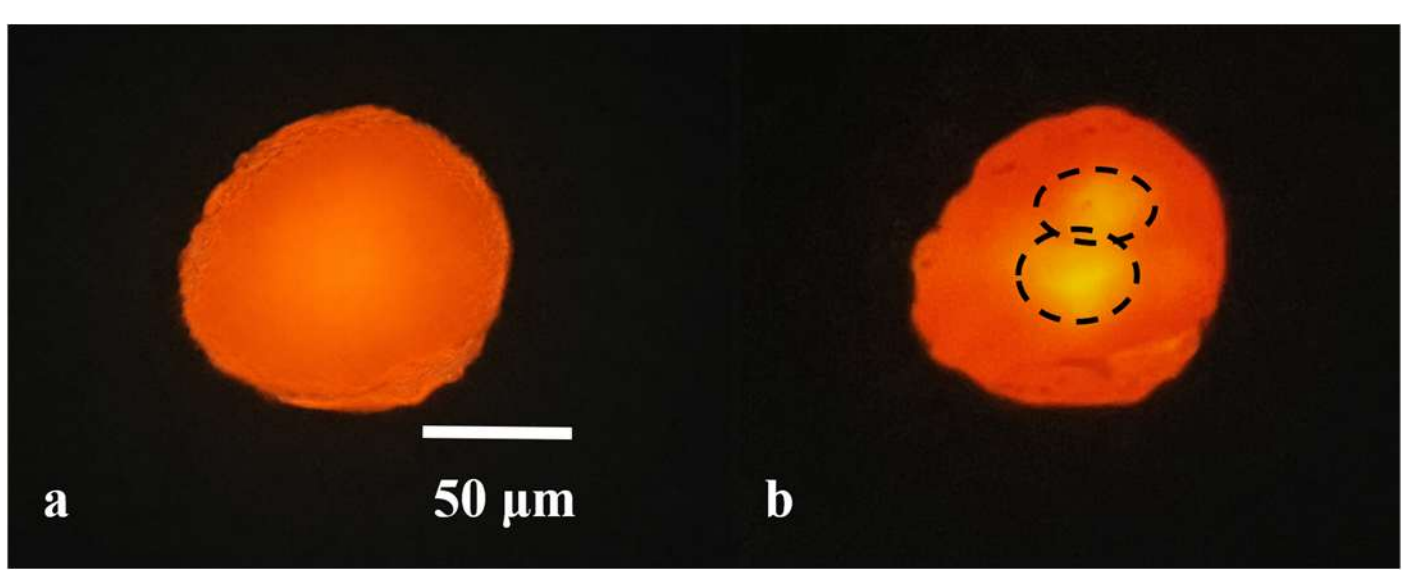

**Figure 3:** Stereomicroscope image of a tholin powder sample before (a) and after (b) the exposure to laser powers exceeding 850 μW under transmitted light illumination. Multiple regions outlined

by black dashed lines exhibiting varying degrees of laser-induced bleaching are visible as lighter-colored and more transparent areas across the sample.

## 4 Implications

Robust constraints on the aggregate elastic properties of tholins provide an important foundation for interpreting future seismic observations acquired by the upcoming Dragonfly mission, enabling improved understanding of the mechanical properties and subsurface layering of Titan's crust (Lorenz et al. 2019). Among the elastic properties determined in this study, the shear modulus is particularly important, since it governs resistance to shear deformation and controls shear-wave propagation within Titan's crust. Variations in shear rigidity are also commonly associated with changes in seismic attenuation and dispersion in materials (Mavko et al., 2009; Carcione, 2014). Consequently, accurate determination of the shear modulus of tholin combined with future Dragonfly seismic data, can also provide important insights into possible physicochemical interactions between Titan's surface organics and subsurface liquid reservoirs, since partial melting, fluid saturation, and chemical alteration reactions commonly produce substantial reductions in shear rigidity accompanied by increased seismic attenuation (Faul et al. 2004; Jackson, 2007; Nimmo & Pappalardo, 2016). Such observations may therefore help identify regions affected by cryovolcanic activity or reactions between surface organic materials and water-rich subsurface ocean-derived materials like clathrates.

In addition, the shear modulus directly affects sediment mobilization thresholds, grain interaction behavior, and the resistance of dune materials to deformation under Titan's low-gravity and dense-atmosphere conditions (Burr et al., 2015; Lapôtre et al., 2016; Yu et al., 2017). The comparatively high elastic stiffness and shear resistance of tholins relative to other Titan-relevant organics support the inference that Titan haze particles play a major role in the formation and long-term stability of Titan's equatorial dune fields. Their high resistance to deformation and high cohesivity suggest haze particle-rich sediments contribute significantly to the mechanical strength, durability, and transport behavior of Titan's dune materials over geological timescales (Yu et al., 2017, 2020, 2021).

The highly consistent elastic properties measured for tholin powder synthesized from a gas mixture of 95% $N_2$ and 5% $CH_4$, and tholin films synthesized from a 90% $N_2$ and 10% $CH_4$ mixture, further suggest that modest variations in the relative abundances of $N_2$ and $CH_4$ during synthesis are unlikely to significantly alter the mechanical properties of tholins. This relative insensitivity of elasticity to small compositional variations implies that tholins produced under a range of Titan-relevant atmospheric conditions may exhibit broadly comparable mechanical behavior. Beyond Titan, tholin-like organic materials are thought to be widespread throughout the outer Solar System. Similar complex refractory organics have been proposed or detected on Pluto, Triton, trans-Neptunian objects (TNOs), Kuiper Belt objects, and on the surfaces of many icy satellites and cometary bodies (Khare et al., 1984; McDonald et al., 1994; Cruikshank et al., 2005, 2019b; Materese et al., 2014). Photochemical and radiation-driven processing of $N_2$-, $CH_4$- and CO-bearing ices under outer solar system conditions can also generate a broad range of tholin-like macromolecular organic materials. Although the detailed chemical compositions of these organics likely vary among planetary environments because of differences in initial volatile inventories, the agreement between the elastic properties of the thin film and powder samples suggests the mechanical behavior of tholin-like materials may be relatively insensitive to variations in sample morphology and small variations in $CH_4/N_2$ precursor ratios. Consequently, the elastic properties determined here for Titan tholins may provide a useful first-order analog for understanding the mechanical properties of complex refractory organics formed in similar conditions to those of Titan. Future work investigating the elastic properties of tholin-like materials under different irradiation environments, temperatures, and precursor compositions can provide insight into how broadly these mechanical properties can be generalized across other outer solar system environments.

This study also demonstrates the effectiveness of Brillouin spectroscopy as a non-destructive and highly sensitive technique for characterizing the elastic properties of organic planetary materials. Unlike conventional mechanical testing methods, Brillouin spectroscopy enables direct determination of acoustic velocities and elastic moduli from extremely small sample volumes without requiring extensive sample preparation or physical contact with the material. This capability is particularly important for fragile, porous, or chemically reactive planetary materials such as tholins, which can be difficult to characterize using traditional mechanical approaches. The successful measurements obtained here further demonstrate that Brillouin spectroscopy can be applied to both thin film and compressed powder samples, highlighting its versatility for studying a broad range of Titan-relevant organic compounds and other complex extraterrestrial materials. Because Brillouin spectroscopy can also be coupled with diamond anvil cell techniques, the method offers significant potential for future investigations of the pressure- and temperature-dependent elastic behavior of planetary organics under simulated outer Solar System environments.

While water- or oxygen-based alteration of haze particles is unlikely under Titan's surface conditions given the absence of free oxygen and negligible water vapor pressure, the observed persistent changes in tholins' elastic properties highlight the sensitivity of tholins to terrestrial environment exposure and provide important guidance for sample handling, storage, and experimental protocols, ensuring reliable and reproducible laboratory measurements. The measured increase in elastic stiffness after approximately one week of air exposure suggests that tholins undergo physicochemical modification during interaction with Earth's atmosphere. These results indicate that the mechanical properties of tholins can evolve significantly during storage and handling under ambient laboratory conditions, potentially introducing systematic uncertainties into experimental measurements if environmental exposure is not carefully controlled. Consequently, tholins and related planetary organic analog materials should ideally be synthesized, stored, and transferred under vacuum or inert-gas environments to minimize atmospheric alteration prior to experiments. More broadly, these findings suggest that environmental exposure history may represent an important factor contributing to variability among previously reported physical property measurements of tholins and other complex planetary organic materials.

# 5 Conclusions

The aggregate elastic properties of tholins were measured using Brillouin spectroscopy, representing the first direct experimental determination of their full set of aggregate elastic properties. The derived $E$ for both samples is consistent with previous nanoindentation studies, supporting the reliability of this approach. Compared to other Titan-relevant organics, tholins exhibit the highest elastic moduli measured to date. Measurements after ~1 week of air exposure reveal persistent stiffening, emphasizing the importance of careful sample handling and storage. Overall, this study demonstrates the effectiveness of Brillouin spectroscopy for characterizing planetary organics and provides new constraints on the mechanical behavior of Titan surface materials relevant for future geophysical investigations.

## Acknowledgments

This study was primarily supported by the Rosie M. and Murray D. Page Geology & Geophysics Endowed Professorship and the Undergraduate Research Program at Texas A&M University (J.S.Z.). Geospectroscopy laboratory instrumentation, M.A.J., W.S. and J.-S.H. were partially supported by NSF-EAR 2240506 (J.S.Z.) and NSF-EAR 2243184 (J.S.Z.). W.S. acknowledges support from the John W. Handin Fellowship at the Center for Tectonophysics, Texas A&M University. A. H., J. R. M.-B., and X. Y. are partially supported by the NASA Cassini Data Analysis Program Grants 80NSSC21K0528 and 80NSSC24K0203, the NASA Planetary Science Early Career Award 80NSSC23K1108, the NASA Solar System Workings Grant 80NSSC24K0888, the NASA Planetary Data Archiving, Restoration and Tools Grant 80NSSC25M7002, the NASA Emerging Worlds Grant 80NSSC26K0080, and the Heising-Simons Foundation grant 2023-3936. A.W. was supported by the NASA MIRO Center for Advanced Measurements in Extreme Environments, Grant 80NSSC19M019.

**Open Research**

All experimental data supporting this study are available in a public data repository at the following link: https://doi.org/10.5281/zenodo.20115094 (Johnson et al. 2026).

**Conflict of Interest Disclosure**

The authors declare there are no conflicts of interest for this manuscript.

Supporting Information for

**Elastic Properties of Titan Tholins Revealed by Brillouin Spectroscopy**

Mark A. Johnson[1], Jose R. Montes-Bojorquez[2], Adis Husić[2], Xinting Yu[2], Wenhao Su[1], Jia-Sen Hu[1], Ryan Blase[3], Edward Patrick[3], Alan Whittington[2], and Jin S. Zhang[1]

[1]Department of Geology and Geophysics, Texas A&M University, College Station, TX, USA.

[2]Department of Physics and Astronomy, University of Texas at San Antonio, San Antonio, TX, USA.

[3]Southwest Research Institute, San Antonio, TX, USA.

## Contents of this file

Figures S1 to S3
Table S1

## Introduction

This supporting information contains the details of the ruby fluorescence measurements (Figure S1), the Fourier Transform Infrared (FTIR) measurements (Figure S2), and a demonstration that the choice of filter applied before peak fitting does not affect the obtained sound velocities (Figure S3 and Table S1).

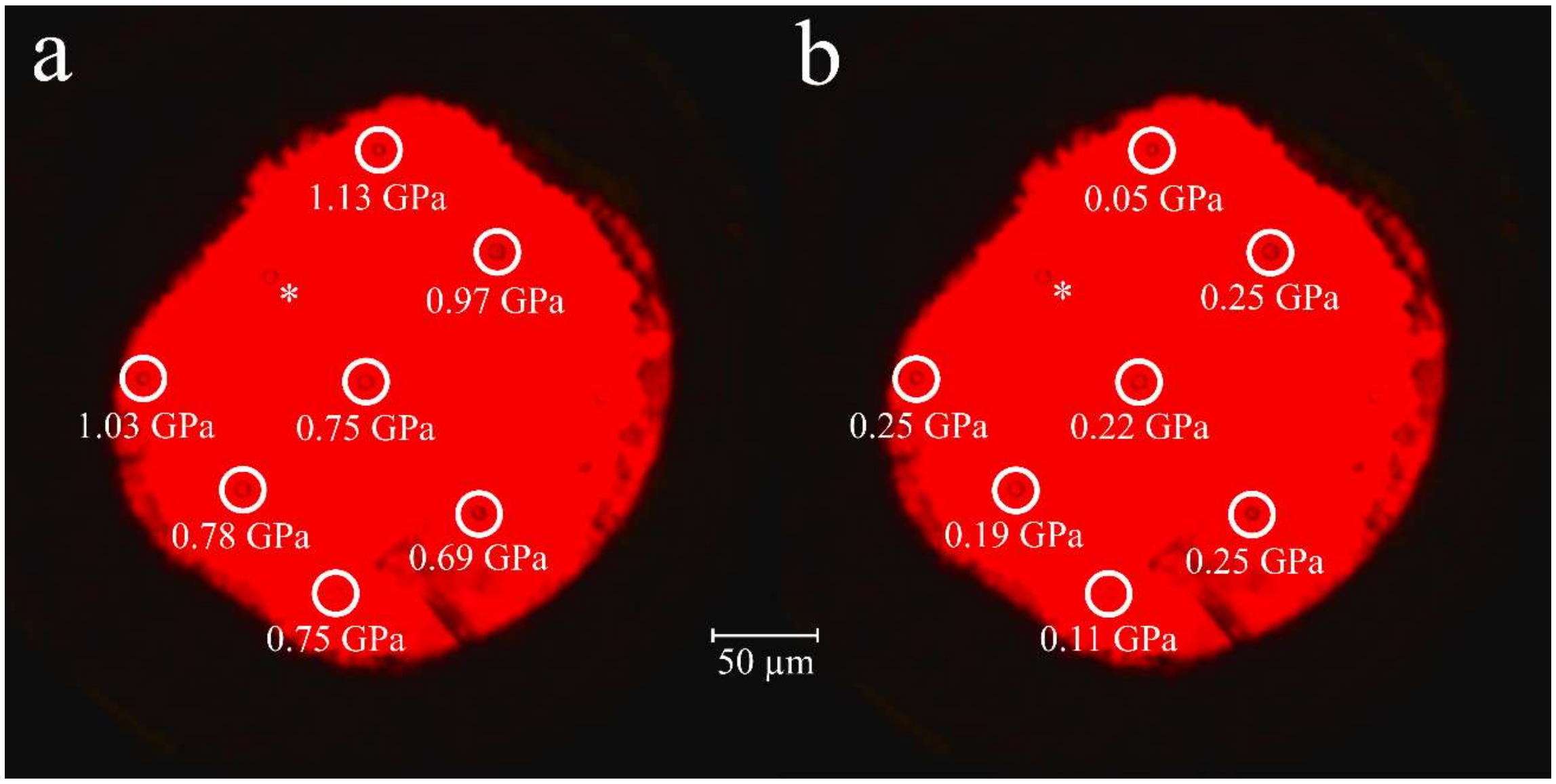

**Figure S1.** Pressures determined by ruby fluorescence measurements at different locations within the sample chamber: (a) during powder compaction, (b) after decompression. No ruby fluorescence spectrum was obtained on the sphere marked with an asterisk, likely due to insufficient Cr doping, resulting in an undetectable fluorescence signal.

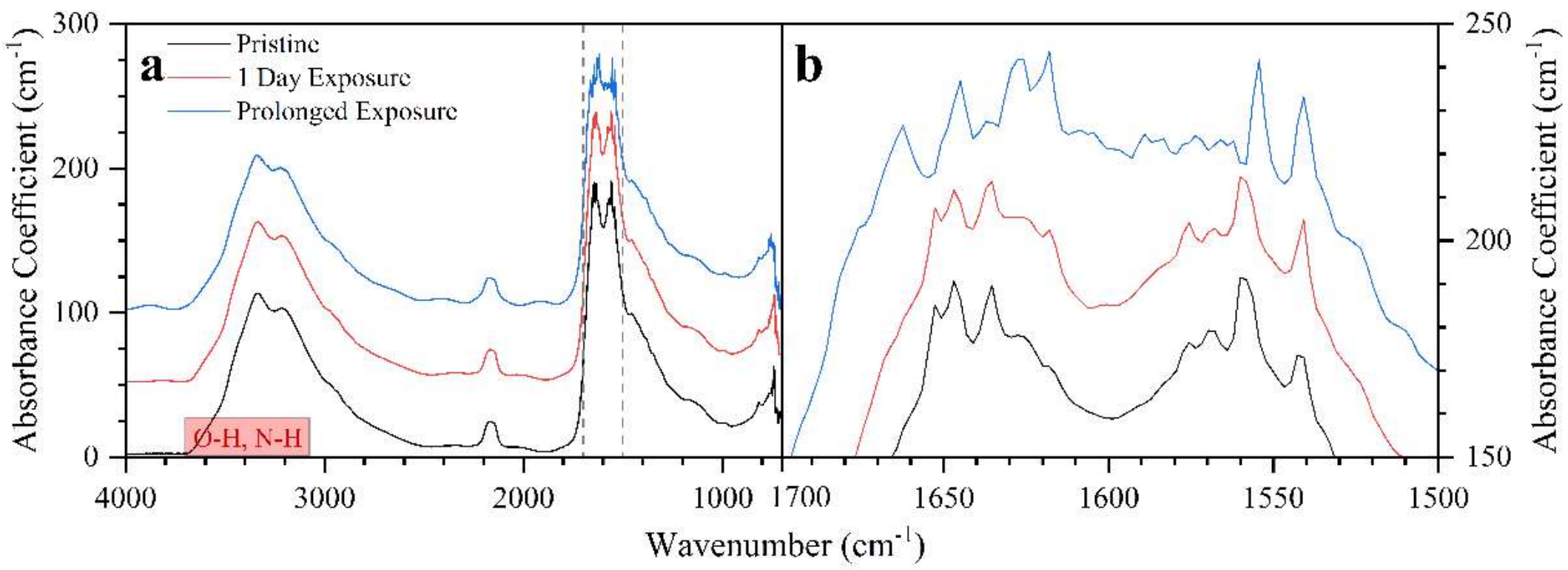

**Figure S2.** Fourier transform infrared (FTIR) absorption coefficient spectra of powder tholin pellets measured before ambient atmospheric exposure, after 1 day of exposure, and after prolonged (>1 week) exposure. Spectra are vertically offset for clarity. (a) Full FTIR spectra, with a constant offset of 50 $cm^{-1}$ in absorption coefficient between successive spectra. (b) Enlarged view of the 1500–1700 $cm^{-1}$ region highlighted by the gray dashed box in (a); the 1-day spectrum is offset by 25 $cm^{-1}$, and the >1-week spectrum is offset by an additional 40 $cm^{-1}$. Spectra were collected using a Bruker Hyperion II FTIR microscope equipped with an MCT-A mid-infrared detector at the Geospectroscopy Laboratory, Texas A&M University. For each measurement, 512 scans were averaged in transmission mode at a spectral resolution of 4 $cm^{-1}$. Reference spectra for background subtraction were collected from an empty location on the $CaF_2$ substrate under otherwise identical measurement conditions. The absorption coefficient was calculated as 2.303*A/d*, where *A* is the absorbance and *d* is the pellet thickness. Measurements were conducted on two pellets: a 15-μm-thick pellet used for the pristine and 1-day exposure measurements and a 20-μm-thick pellet used for the prolonged-exposure measurement.

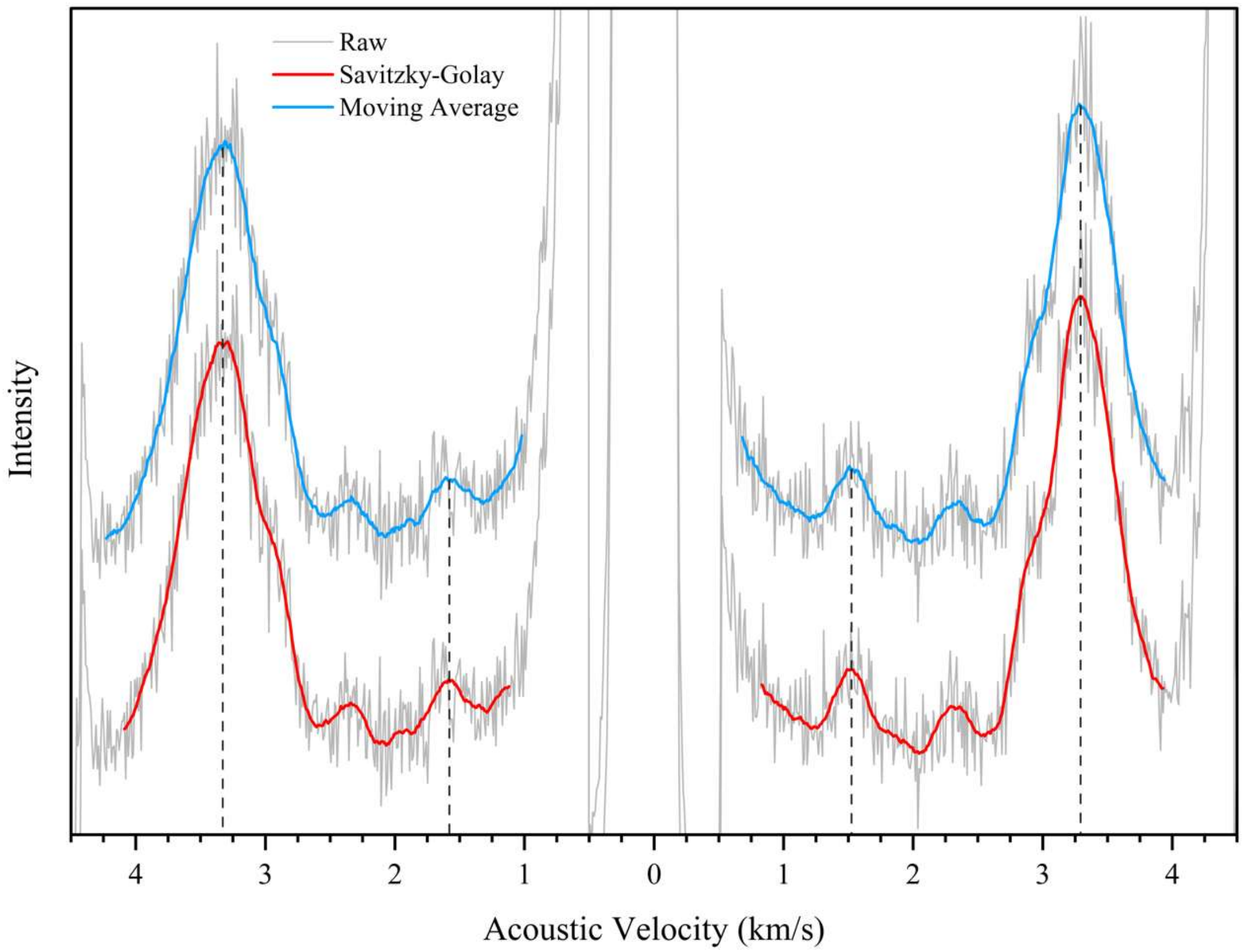

**Figure S3.** Comparison of the Brillouin spectrum shown in Figure 1c after application of a low-pass Savitzky–Golay (S-G) filter and a low-pass moving-average (MA) filter with comparable smoothing performance. Both filters produce nearly identical peak positions and widths, indicating that the choice of smoothing filter does not affect the fitted elastic properties.

| Velocity interval (km/s) | S-G fitting error (counts) | MA fitting error (counts) |
|---|---|---|
| (-4, -2.5) | 17.00 | 17.14 |
| (-2, -1.25) | 13.84 | 13.77 |
| (1.25, 2) | 13.14 | 13.44 |
| (2.5, 4) | 16.29 | 16.83 |

**Table S1.** Fitting errors for the S-G filter and MA filter in Figure S3